\documentclass[sigconf]{acmart}

\copyrightyear{2023} 
\acmYear{2023} 
\setcopyright{acmlicensed}
\acmConference[SIGIR '23]{Proceedings of the 46th International ACM SIGIR Conference on Research and Development in Information Retrieval}{July 23--27, 2023}{Taipei, Taiwan}
\acmBooktitle{Proceedings of the 46th International ACM SIGIR Conference on Research and Development in Information Retrieval (SIGIR '23), July 23--27, 2023, Taipei, Taiwan}
\acmPrice{15.00}
\acmDOI{10.1145/3539618.3591844}
\acmISBN{978-1-4503-9408-6/23/07}
\usepackage{booktabs}
\usepackage{multirow}
\usepackage{graphicx}
\usepackage[title]{appendix}

\usepackage[section]{placeins}
\begin{document}

\title{Enhancing Dynamic Image Advertising with Vision-Language Pre-training}
\vspace{-2mm}


\author{Zhoufutu Wen\footnotemark[1]}
\affiliation{%
 \institution{Baidu Search Ads, Baidu Inc.}
  \city{Beijing}
 \country{China}}
 \email{wenzhoufutu01@baidu.com}

\author{Xinyu Zhao\footnotemark[1]\footnotemark[2]}
\affiliation{%
 \institution{Peking University}
   \city{Beijing}
  \country{China}}
\email{xinyuzhao@stu.pku.edu.cn}

\author{Zhipeng Jin}
\affiliation{%
 \institution{Baidu Search Ads, Baidu Inc.}
  \city{Beijing}
 \country{China}}
\email{jinzhipeng@baidu.com}
\vspace{-2mm}
 
\author{Yi Yang\footnotemark[3]}
\affiliation{%
 \institution{Baidu Search Ads, Baidu Inc.}
  \city{Beijing}
 \country{China}}
\email{yangyi15@baidu.com}

\author{Wei Jia}
\affiliation{%
 \institution{Baidu Search Ads, Baidu Inc.}
  \city{Beijing}
 \country{China}}
\email{jiawei04@baidu.com}

 \author{Xiaodong Chen}
\affiliation{%
 \institution{Baidu Search Ads, Baidu Inc.}
  \city{Beijing}
 \country{China}}
\email{chenxiaodong@baidu.com}
\vspace{-2mm}

\author{Shuanglong Li}
\affiliation{%
 \institution{Baidu Search Ads, Baidu Inc.}
  \city{Beijing}
 \country{China}}
\email{lishuanglong@baidu.com}

\author{Lin Liu}
\affiliation{%
 \institution{Baidu Search Ads, Baidu Inc.}
  \city{Beijing}
 \country{China}}
\email{liulin03@baidu.com}

\renewcommand{\shortauthors}{Zhoufutu Wen et al.}

\def \authors{Zhoufutu Wen, Xinyu Zhao, Zhipeng Jin, Yi Yang, Wei Jia, Xiaodong Chen, Shuanglong Li, and Lin Liu}
% 这是ACM Reference Format位置显示的作者名

\begin{abstract}
In the multimedia era, image is an effective medium in search advertising. Dynamic Image Advertising (DIA), a system that matches queries with ad images and generates multimodal ads, is introduced to improve user experience and ad revenue. The core of DIA is a query-image matching module performing ad image retrieval and relevance modeling. Current query-image matching suffers from limited and inconsistent data, and insufficient cross-modal interaction. Also, the separate optimization of retrieval and relevance models affects overall performance. To address this issue, we propose a vision-language framework consisting of two parts. First, we train a base model on large-scale image-text pairs to learn general multimodal representation. Then, we fine-tune the base model on advertising business data, unifying relevance modeling and retrieval through multi-objective learning. Our framework has been implemented in Baidu search advertising system ``Phoneix Nest''. Online evaluation shows that it improves cost per mille (CPM) and click-through rate (CTR) by 1.04\% and 1.865\%.
\vspace{-2mm}
\end{abstract}

\vspace{-2mm}
\vspace{-2mm}


\begin{teaserfigure}
  \includegraphics[width=\linewidth]{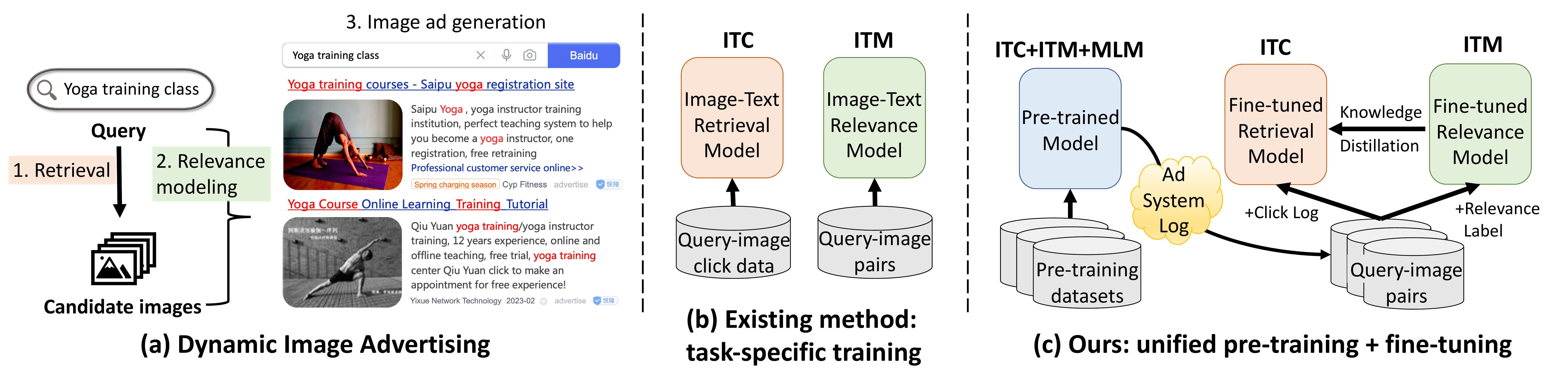}
  \caption{Dynamic Image Advertising illustration: (a) First, candidate images are retrieved based on their distance from query. Then, the images are re-ranked according to query-image relevance. (b) Previous solution is training task-specific models separately. (c) We adopt vision-language pre-training and fine-tuning to unify retrieval and relevance models.}
  \label{fig:teaser}
\end{teaserfigure}

\maketitle

\renewcommand{\thefootnote}{\fnsymbol{footnote}} %将脚注符号设置为fnsymbol类型，即特殊符号表示
\footnotetext[1]{Equal contribution. $^{\ddag}$Corresponding author.} 
\footnotetext[2]{Work done during an internship at Baidu Search Ads.}

\section{Introduction}
In the era of rich media, multimodal advertisement (ad) especially image ad is becoming a popular form in search advertising. Baidu, as one of the leading search companies, introduces Dynamic Image Advertising (DIA), a system that pairs queries with ad images and generating multimodal ads, see Figure \ref{fig:teaser}. The core of the system is a query-image matching module. Given a query, it first retrieves candidate images via approximate nearest neighbor (ANN) search, and then computes query-image similarity to rerank images for subsequent procedures such as click-through rate (CTR) prediction. A powerful query-image matching module not only improve user experience, but also increase ad clicks and revenue for advertisers and search engine. However, two aspects make query-image matching a challenging task: the multimodal and multi-objective problems.

Regarding the multimodal feature, the queries and ad images are in two drastically different feature spaces, which need to be reconciled through cross-modal representation learning. Previous solution is to train models specifically for relevance and retrieval task, affecting overall performance. And the small models suffering from limited multimodal data and insufficient cross-modal interaction. In recent years, Vision-Language Pretraining (VLP) \cite{vilt} achieves good performance in cross-modal tasks. Its main idea is self-supervised learning on large-scale unlabeled multimodal data to generate latent representations for downstream tasks. VLP is compatible with query-image matching, because there is a large number of query-image pairs available, and learning cross-modal representation helps measuring query-image similarity. 

As for the multi-objective aspect, due to the commercial nature of advertising, the system goal is to generate ads that are not only relevant to the query, but also higher in CTR. Previous matching module consists of two single-objective models: a retrieval model to recall high-CTR ad images, and a relevance model to rank query-image similarity. Single-objective training may result in presenting ads that are either not relevant or not attractive. Also, developing separate models increases training and maintenance cost.

Given the existing problems, in this paper, we propose a vision-language framework for query-image matching, composing of two parts. First, to learn general multimodal representations, we train a base model on large-scale multimodal datasets. Then, we further train the base model to build a relevance model and a multitask retrieval model. In the online and offline evaluation, our framework shows good representing ability and profitability. The contributions of this paper are as follows: 

\noindent $\bullet$ To improve image matching for diverse queries, on multi-domain multimodal datasets of 20B image-text pairs, we train a base model, incorporating dual encoders optimized by momentum contrastive loss (ITC), and a fusion encoder optimized by masked language modeling (MLM) and image-text matching (ITM). With the base model, image diversity ratio increases by over 3\% and irrelevant case ratio on long-tail queries declines by 10\%. 

\noindent $\bullet$ To transfer the multimodal representation learning ability to advertising scenario, we further train the base model on query-image pairs with relevance labels. Our relevance model outperforms previous one in AUC by over 4\%.

\noindent $\bullet$ To improve ad quality and profitability simultaneously, we unify knowledge distillation from the relevance model and further contrastive learning on ad click data. After launching the multitask retrieval model, CPM and CTR of our search advertising system grow by 1.04\% and 1.865\%
\vspace{-2mm}

\section{Related Work}

Inspired by the success of pre-training in NLP, VLP has progressed rapidly in recent years and has developed several sturctures. Some models consist only of encoders. According to how the two modalities interact, encoder models can be divided into two types. One deeply fuses vision and language embeddings through transformer layers, called fusion encoder, which is suitable for vision-language understanding tasks \cite{uniter, oscar, vilt}. Another type, called dual encoder model, first encodes vision and language embeddings independently and then aligns them by metric learning, which makes it more efficient in vision-language retrieval tasks \cite{clip, align}. Recently, there are studies combining different types of encoders and even decoders to exploit their strengths for different tasks \cite{albef, simvlm, coca, blip, vlmo}. Besides, image feature extraction has been the computational bottleneck of VLP. As vision transformer \cite{vit} has been proposed, many studies use it as vision backbone in VLP instead of object detectors and convolutional networks (CNN), which enhances model scalability and training efficiency \cite{vilt, albef, vlmo}. 

Ad relevance modeling, an essential part in search advertising, is to understand search intent and then match relevant and profitable ads. Previous studies introduce commercial metrics such as CPM and CTR as additional optimization goals besides query-ad similarity \cite{mobius, uniretriever}.
Recently, VLP has also been applied to multimodal search and search advertising. Trained on multimodal fashion data, FashionBERT and Kaleido-BERT are proposed to solve ITR in the e-commerce industry \cite{fashionbert, kaleidobert}. AdsCVLR models the relevance among query, ad image and ad text \cite{adscvlr}. Baidu, as a leading search engine provider in China, has also applied VLP to improve the online advertising platform and released a series of works \cite{han, tira, boostcp}. However, existing works focus on relevance modeling ability and train multimodal models with only domain data. There is still room to better utilize general VLP model and data, and balance ad relevance and CTR in advertising scenarios. 
\vspace{-2mm}

\section{Methods}
\begin{figure*}[h]
  \centering
  \includegraphics[width=\linewidth]{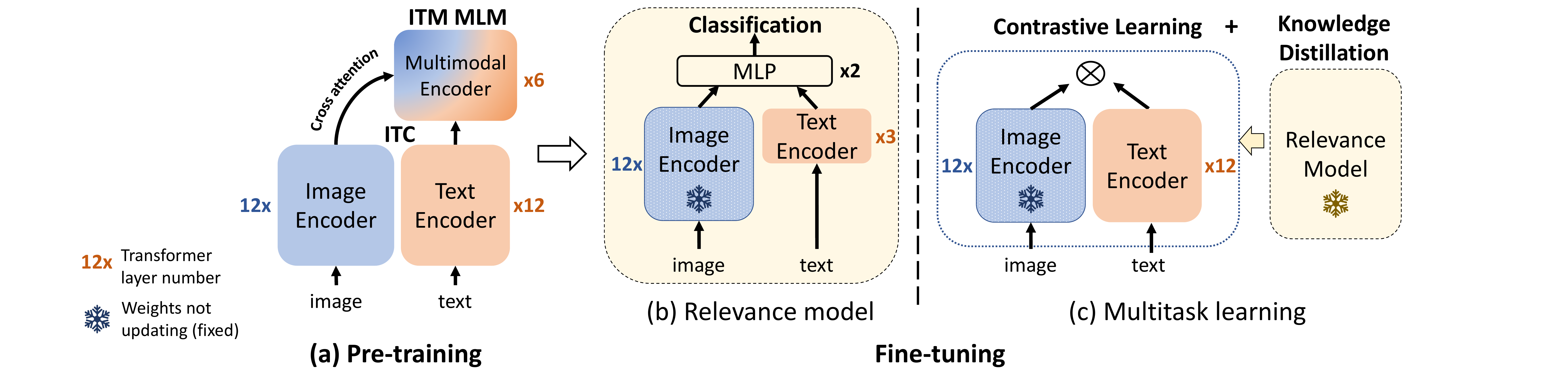}
  \caption{Overview of our vision-language framework. (a) We jointly pre-train two single-modal encoders with contrastive loss, and a multimodal encoder that fuses modalities through image-to-text cross attention with masked language modeling and image-text matching loss. Then, we fine-tune the single-modal encoders on two tasks, with the image part fixed (not updating parameters): (b) classification of relevant query-image pairs; (c) computing query-image cosine similarity to discriminate clicked pairs and fit relevance teacher model output simultaneously.}
  \label{model}
  \vspace{-2mm}
\end{figure*}

With VLP, we can improve query-image matching for long-tail, low-frequency queries. And we need a unified model to balance ad quality and profitability. Therefore, we propose a vision-language framework for image advertising. As shown in Figure \ref{model}, it follows the ``pre-training and then fine-tuning'' paradigm, composing of three models: pre-trained base model, fine-tuned relevance model, and multitask retrieval model.

\vspace{-2mm}

\subsection{Pre-training}
The pre-trained model consists of two single-modal encoders and a multimodal fusion encoder. The vision and language encoders are trained to align two modalities, while the fusion encoder helps capture fine-grained cross-modal interaction. We use ViT-B/16 \cite{vit} and RoBERTa$_{base}$ \cite{roberta} as the vision and language encoders, processing embeddings of each modality separately with self-attention. The single-modal encoders are trained on multi-view Image-Text Contrastive learning (ITC) \cite{ernievil}. The contrastive loss, for which we use InfoNCE \cite{infonce}, aims to maximize the contrast between the similarity of positive pairs and that of negative pairs. Our multi-view ITC loss combines bidirectional cross-modal contrastive losses and two single-modal contrastive losses. The cross-modal contrastive learning helps align the text and image feature spaces, while the single-modal loss is to enhance the ability to distinguish the semantics within each modality. Following \cite{moco}, we enhance negative sampling in ITC with momentum encoders, which are the exponential moving average version of single-modal encoders.

Then, the output of the language encoder is fed into the fusion encoder, implemented with RoBERTa$_{base}$. It computes cross-modal attention that receives visual features as key and value inputs, producing multimodal language representation conditioned on the vision modality. The fusion encoder is trained on multimodal masked language modeling (MLM), and image-text matching (ITM). Our multimodal MLM is to predict masked language tokens based on both language context and image. Similar to BERT \cite{bert}, the objective function of MLM is the cross-entropy loss between the prediction and the ground truth. ITM is to classify whether an image matches with a text (positive) or not (negative). The objective function of ITM is the cross-entropy loss between the classifier output and the ground truth label. We improve ITM with hard negative mining following \cite{albef}, which encourages model to learn from meaningful negative pairs with high similarity. 
% Details of the pre-training loss are explained in Appendix \ref{appendix_b}. 

\vspace{-2mm}

\subsection{Fine-tuning}

To adapt our base model with general representing ability to different roles in the query-image matching module, we design two fine-tuning strategies, obtaining a relevance model and a multitask retrieval model. In fine-tuning, we stop updating the parameters of the vision encoder, to facilitate computation and prevent overfitting.

\noindent $\bullet$ \textbf{Relevance Model.} To ensure ad quality, we expect high-relevant images to be selected. But a large part of pre-training data are noisy image-text pairs crawled from the Web that are not necessarily relevant. Besides, ad queries are shorter and more abstract than texts in general datasets. Therefore, we further train our base model on business data with relevance labels, transferring its representing ability to advertising domain data.

The relevance model is fine-tuned on the dual encoders of our base model. Before fine-tuning, we use the base model to temporarily replace the retrieval model generating candidate image embeddings. In fine-tuning, candidate images and queries are fed into the relevance model. We place a MultiLayer Perceptron (MLP) on top of the dual encoders as classifier, which computes two-dimensional logits. We minimize the cross-entropy loss between the logits and relevance labels.

\noindent $\bullet$ \textbf{Multitask Retrieval Model.} In query-image matching module, the input candidate images of the relevance model are determined by the retrieval model. To ensure ad quality and mine the commercial value of image ad and, we need to preserve the knowledge of query-image relevance, while encouraging high-CTR ad images to be favored by the retrieval model. Thus, we design a retrieval model with two objectives: modeling image commercial value on click data while learning query-ad relevance from the relevance teacher model. 

The multitask retrieval model is also fine-tuned from the dual encoders of our base model. The dual encoders learn a cosine similarity between query and image embeddings, jointly optimized by contrastive learning and knowledge distillation. For contrastive learning, we minimize the bidirectional cross-modal contrastive loss, to distinguish clicked query-image pairs (positive) from the others (negative). Different from the ITC in base model, we use in-batch negative sampling to facilitate computation. For knowledge distillation, we minimize the Euclidean distance between the cosine similarity and the classifier output of relevance model.

\vspace{-3mm}

\section{Experiments}

\noindent $\bullet$ \textbf{Datasets.} For pre-training base model, we gather 20B image-text pairs, including general image-text pairs, and advertising domain data from Baidu DIA system which are query-image pairs clicked more than once from January 2021 to July 2022. During pre-training, we perform data augmentation with RandAugment \cite{randaug}.
The dataset for fine-tuning relevance model is about 500K query-image pairs from the DIA system that are annotated with relevance degree $\in \left \{0, 1, 2\right \} $, where 0 is negative and 1 or 2 is positive.
The data for multitask fine-tuning consists of 1.3B query-image pairs with at least 2 clicks in the past six months from Baidu DIA system, including about 2K unique images.

\noindent $\bullet$ \textbf{Evaluation of the base model.} To evaluate the representing ability of our base model (ours$_{base}$), we compare it with the base-size Chinese CLIP (CN-CLIP), the SOTA model in Chinese cross-modal retrieval \cite{cnclip}. 

We experiment on both Chinese public benchmark datasets and our business datasets. The public datasets include MSCOCO-CN \cite{cococn}, Flickr30k-CN \cite{flickr30kcn}, Wukong \cite{wukong}. Three private datasets are from different domains: search, advertising, and e-commerce. We evaluate model performance by text-to-image Recall@$K$, $K\in \left \{ 1, 5,10 \right \}$, which measures whether the ground truth is included in the top K results.


\begin{table}[]
\caption{Zero-shot performance on text-to-image retriveal on public and private datasets.}
\label{pretrain_result}
\resizebox{\columnwidth}{!}{%
\begin{tabular}{@{}cccccccccc@{}}
\toprule
\multirow{2}{*}{Method} & \multicolumn{3}{c}{Wukong} & \multicolumn{3}{c}{MSCOCO-CN} & \multicolumn{3}{c}{Flickr30-CN} \\ \cmidrule(l){2-10} 
 & R@1 & R@5 & R@10 & R@1 & R@5 & R@10 & R@1 & R@5 & R@10 \\ \midrule
CN-CLIP & 45.6 & 72.4 & 79.8 & \textbf{62.2} & \textbf{86.6} & \textbf{94.9} & \textbf{62.7} & \textbf{86.9} & \textbf{92.8} \\
ours$_{base}$ & \textbf{56.3} & \textbf{82.9} & \textbf{88.0} & 51.0 & 80.8 & 91.2 & 45.8 & 75.2 & 84.2 \\ \midrule
\multirow{2}{*}{Method} & \multicolumn{3}{c}{search} & \multicolumn{3}{c}{advertising} & \multicolumn{3}{c}{e-commerce} \\ \cmidrule(l){2-10} 
 & R@1 & R@5 & R@10 & R@1 & R@5 & R@10 & R@1 & R@5 & R@10 \\ \midrule
CN-CLIP & 23.6 & 48.9 & 59.0 & 8.0 & 22.1 & 30.6 & \textbf{58.9} & 79.7 & 84.8 \\
ours$_{base}$ & \textbf{36.1} & \textbf{67.4} & \textbf{76.7} & \textbf{9.8} & \textbf{27.0} & \textbf{37.4} & 58.0 & \textbf{80.0} & \textbf{85.5} \\ \bottomrule
\end{tabular}%
}
\vspace{-2mm}
\end{table}

As shown in Table \ref{pretrain_result}, the overall performance of our base model (ours) is comparable to CN-CLIP in public datasets, with a distinct advantage in business domain datasets, proving that it is powerful in commercial scenarios, while maintaining expressiveness for general data. The results on public datasets also show the influence of data distribution in multimodal pre-training. Pre-trained on translated descriptive image-text pairs (Visual Genome \cite{vg} and MSCOCO), CN-CLIP performs better on MSCOCO-CN and Flickr30K-CN that share the same data source, while our base model gains advantage on Wukong based on original noisy Chinese web content as our pre-training data.

\begin{table}[]
\caption{Experiment result of image retrieval between task-specifically trained model and pre-trained model.}
\label{img_recall}
\begin{tabular}{@{}ccc@{}}
\toprule
Method & Diversity Ratio & Irrelevant Ratio \\ \midrule
previous$_{retrieval}$ & 6.11 & 32.67\\
ours$_{base}$ & \textbf{9.40} & \textbf{21.78} \\ \bottomrule
\end{tabular}
\vspace{-2mm}
\end{table}

To evaluate model performance on image retrieval, we index a dataset of 860K images by our base model (ours$_{base}$) and the previous retrieval model (previous$_{retrieval}$). Then, we randomly select 10K queries, conducting ANN search to retrieve their top-15 similar images from each index, to examine the ratio of retrieved images to all images (Diversity Ratio). In addition, we perform query-image matching on 1K long-tail queries on the two indexes, and invite annotators to label the query-image relevance (0 for irrelevant, 1 for relevant, and 2 for highly relevant). We record the ratio of pairs labeled as irrelevant (Irrelevant Ratio). As shown in Tabel \ref{img_recall}, our base model surpasses previous task-specific model in both image retrieval on general queries and long-tail queries. In addition, we visualize the query-image matching results with our proposed framework in Appendix \ref{appendix_a}. 

\begin{table}[]
\caption{Comparison between relevance models.}
\label{auc}
\begin{tabular}{@{}cc@{}}
\toprule
Method & AUC \\ \midrule
previous$_{relevance}$ & 82.00 \\
ours$_{relevance}$ & \textbf{86.62} \\ \bottomrule
\end{tabular}%
\vspace{-2mm}
\end{table}

\noindent $\bullet$ \textbf{Evaluation of the relevance model.} To evaluate our relevance model (ours$_{relevance}$), we compare it with the previous relevance model (previous$_{relevance}$) \cite{mixbert}. They are tested on 10K query-image pairs, with AUC (area under the curve) as the metric. the previous model is trained to match a triple <query, image description, image>. While our relevance model separately encodes text query and visual content all by transformer. As shown in Table \ref{auc}, the better performance of our relevance model demonstrates that it is possible to directly model vision-language interaction without image description as a medium. We also perform human evaluation on the relevance of query-image pairs showed online, where our relevance model outperforms previous one by 7\%.

\begin{table}[]
\caption{Ablation study of multitask fine-tuning.}
\label{recall_rel}
\begin{tabular}{@{}ccc@{}}
\toprule
Method & Recall@10 & Relscore@10 \\ \midrule
base & 75.3 & 78.5 \\
retrieval w/o KD & 92.8 & 77.7 \\
retrieval w/ KD & \textbf{94.1} & \textbf{79.5} \\ \bottomrule
\end{tabular}%
\vspace{-2mm}
\end{table}

\noindent $\bullet$ \textbf{Evaluation of the multitask retrieval model.} For ablation study, we compare our retrieval model fine-tuned with additional relevance knowledge distillation (retrieval w/ KD) to our base model (base) and a retrieval model fine-tuned with contrastive learning only (retrieval w/o KD). They are tested on 10K query-image click data. The evaluation metrics are Recall@10 and Relevance Degree (Rel@10), which is the average relevance score for the top 10 results. As shown in Table \ref{recall_rel}, the single-task retrieval model improves the recall of clicked ad images, but affects the relevance. With multitask fine-tuning, clicked ad recall and image relevance increase simultaneously.

\begin{table}[]
\caption{Online A/B testing result.}
\label{online_test}
\begin{tabular}{@{}cccc@{}}
\toprule
Method & CPM & CTR & P97 Latency \\ \midrule
ours$_{retrieval}$ & +1.04\% & +1.865\% & +0.05\% \\ \bottomrule
\end{tabular}%
\vspace{-2mm}
\end{table}
\noindent $\bullet$ \textbf{Online Experiments.} We compare the CPM, CTR, and 97th percentile tail latency (P97 latency) before and after lauching the proposed framework in Baidu search advertising system for 15 days. As shown in Table \ref{online_test}, the CPM and CTR increase by 1.04\% and 1.865\% respectively, which is a substantial gain considering it is observed on the system main traffic. Given that the new query-image matching module is similar to the previous one in terms of computational cost and model size, the P97 latency shows no discernible increase, verifying our framework improves ad revenue without sacrificing system response time.

\vspace{-2mm}

\section{Conclusion}
In the multimedia era, image ad become a powerful form of search advertising. We propose a vision-language framework for query-image matching in image advertising. We first train a base model with powerful general representing ability, and then fine-tune it on business data to adapt to image advertising scenarios. Furthermore, we jointly optimize the base model to balance ad relevance and CTR. Offline and online experiments demonstrate its profitability and effectiveness in improving the quality of image ads.

\section*{Company Portrait}
Baidu is a leading search engine platform and tech company in China with a strong AI foundation, encompassing AI infrastructure, core AI capabilities, as well as an open AI platform. Baidu provides online marketing services and non-marketing value added services, as well as products and services from new AI initiatives.

\section*{Presenter Bio}
Zhoufutu Wen is an algorithm engineer at Baidu Search Ads. His research interests include multimodal content understanding, cross-modal information retrieval. Wen currently works on the construction of vision-language big model and its applications.
\bibliographystyle{ACM-Reference-Format}
\balance
\bibliography{vican}

\appendix

\section{Case Study}\label{appendix_a}

\setcounter{figure}{0}
\renewcommand{\thefigure}{A\arabic{figure}}

\begin{figure}
  \centering
  \includegraphics[width=0.8\linewidth]{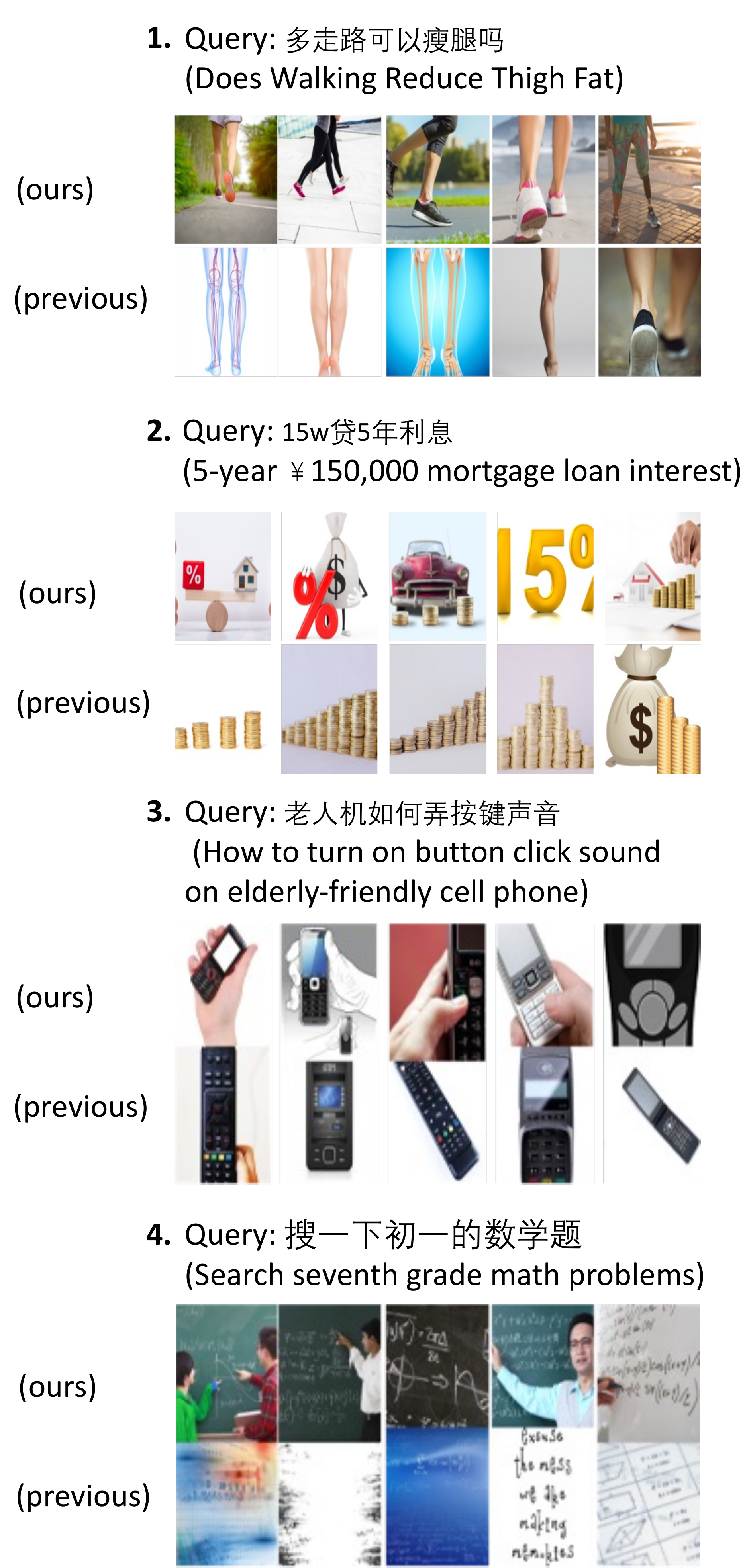}
  \caption{Top-5 matched images for example queries. Queries are in Chinese and translated to English for reading convenience. (ours) denotes images retrieved by our proposed framework. (previous) denotes image retrieved by previous query-image matching module.}\label{fig:demo}
\end{figure}

We compare the query-to-image matching between our framework and previous method in Figure \ref{fig:demo}. For each query, we show the top-5 matched images. The results of our framework are more relevant to user search intent. For example, for the fourth query to search math problems, previous module includes a image of English words and some blurred images in the results, while our method retrieves clearer images about math equations. In the image matched for the third query, the previous module misidentifies the remote control as a cell phone. And our framework outperform previous one in terms of image diversity taking the second query as example, which can possibly improve image ad attractiveness when displaying several similar ads. Examining the detail of queries and images, it can be found out that our framework trained with close vision-language interaction is able to capture fine-grained information from queries and retrieve corresponding images. For instance, for the first query, the images our framework match not only contain the information about ``thigh'' as previous module does, but also capture the information about ``walking''.

\end{document}